\documentclass[prb,a4,twocolumn
                    ]{revtex4}
\usepackage{epsfig}

\def\e{\begin{equation}}
\def\f{\end{equation}}
\def\ea{\begin{eqnarray}}
\def\fa{\end{eqnarray}}
\def\=#1{\overline{\overline{#1}}}
\def\_#1{{\bf #1}}
\def\o{\omega}
\def\E{\epsilon}
\def\M{\mu}
\def\.{\cdot}
\def\x{\times}

\def\l#1{\label{eq:#1}}
\def\r#1{(\ref{eq:#1})}

\def\ds{\displaystyle}

\def\l#1{\label{eq:#1}}
\def\r#1{(\ref{eq:#1})}

\renewcommand{\Re}{\mathop{\rm Re}\nolimits}

\begin{document}

\title{Near-field enhancement and imaging in double
planar polariton-resonant structures}

\author{Stanislav Maslovski}

\author{Sergei Tretyakov}

\author{Pekka Alitalo}

\affiliation{Radio Laboratory / SMARAD, Helsinki University of Technology\\
P.O. Box 3000, FIN-02015 HUT, Finland\\
{\rm E-mails: stanislav.maslovski@hut.fi, sergei.tretyakov@hut.fi, pekka.alitalo@hut.fi}}

\date{\today}

\begin{abstract}
It is shown that a system of two coupled planar material sheets
possessing surface mode (polariton) resonances can be used
for the purpose of evanescent field restoration and, thus, for the sub-wavelength near-field imaging.
The sheets are placed in free space so that they are parallel and
separated by a certain distance. Due to interaction of the
resonating surface modes (polaritons) of the sheets an exponential
growth in the amplitude of an evanescent plane wave coming through
the system can be achieved. This effect was predicted earlier for
backward-wave (double-negative or Veselago) slab lenses. The alternative system considered here is proved to be realizable at microwaves by grids or arrays of resonant particles. The necessary electromagnetic properties of the resonating grids and the particles are investigated and established. Theoretical
results are supported by microwave experiments that demonstrate
amplification of evanescent modes.

\end{abstract}

\maketitle

\section{Introduction}

Negative refraction and sub-wavelength resolution in slab lenses
based on backward-wave (also called {\em double-negative} or {\em Veselago}\cite{Veselago})
materials are the topics that have been widely discussed in the
recent literature. The discussion started soon after Pendry had
published his article\cite{Pendry} claiming a possibility to overcome the
$\lambda/2$ optical resolution limit using a slab of Veselago
material with the relative parameters $\epsilon_r=-1$ and $\mu_r=-1$. The result came out of the
discovery that a Veselago slab\cite{Veselago} can ``amplify'' exponentially
decaying evanescent modes of a source field. The slab restores
the amplitudes of these modes in the image plane recovering the
fine details of a source.

It was no wonder that many scientists, especially those from the
optical community, found this idea of ``amplification''
difficult to accept.\cite{Garcia}  Indeed, if one assumes an exponential growth of
the field amplitude in a slab of a finite thickness, it seems that
increasing the slab thickness the amplitude of the outgoing field can
be made arbitrarily high. However, an accurate analysis shows that if
the slab relative permittivity and permeability are not exact $-1$
(due to inevitable losses and dispersion) that cannot happen.
Indeed, for a slab of thickness $d$ of Veselago material
characterized by $\E_r$, $\M_r$ the slab transmission coefficient
can be found using the standard procedure of expressing the slab
fields in terms of two oppositely propagating (decaying) waves and
finding the unknown wave amplitudes by solving a system of four
equations coming from the boundary conditions on the slab
interfaces. For excitation by an evanescent plane wave with the
tangential component of the propagation factor $k_{\rm t}$ we have
\e T = {2\over 2\cosh(\alpha d) + (\gamma +
1/\gamma)\sinh(\alpha d)} \l{PendryT} \f
Here $\alpha =
\sqrt{k_{\rm t}^2-\E_r\M_r k_0^2}$ is the decay factor in
the slab, $\alpha_0=\sqrt{k_{\rm t}^2-k_0^2}$ is the same for free
space, $\gamma = {\ds\alpha_0\M_r\over\ds\alpha}$ for TE
incidence or $\gamma = {\ds \alpha\over\ds\alpha_0\E_r}$ for TM
case, and $k_0 = \o\sqrt{\E_0\M_0}$. Assuming $\E_r=\M_r=-1$, we see that $\gamma = -1$ and $T =
\exp(\alpha_0 d)$ as in the ideal Pendry's case. However, if the permittivity
and (or) permeability differ from that very value, then $\gamma
\neq -1$ and $|\gamma + 1/\gamma| > 2$, resulting in domination of
the growing exponent in the denominator of \r{PendryT} when the slab
thickness and (or) the incidence field spatial frequency become
large enough. We see that the region where the evanescent fields
are indeed amplified in a Veselago slab is limited by several
factors. Some of them are even inevitable in any realistic
material, e.g., losses and spatial dispersion.\cite{Pavel}

For the following, it is important to understand what main phenomena
lead to amplification of evanescent modes in
the Veselago slab. For this purpose, we will shortly review the
plane wave incidence problem for an interface of free space and a
half-space filled by a Veselago material. In the material we will
look for a solution which, as usually, is either exponentially decaying (if the
transmitted wave is evanescent) or transmitting energy {\it from}
the interface (if the wave is propagating).

If $A$, $B$, and $C$ denote the complex amplitudes of the
incident, transmitted, and reflected wave electric field
tangential components, respectively, then using the interface
boundary conditions we can write:
\e \begin{array}{ccc}
A + C &=& B \\[1mm]
\ds {A - C\over\eta_0} &=& \ds {B\over\eta}\\
\end{array} \l{abc}
\f
Here $\eta_0$ and $\eta$ are the wave impedances that connect the
tangential components of electric and magnetic fields in free
space and in the medium, respectively. The solution of \r{abc} is,
obviously,
\e C = {\eta - \eta_0\over \eta + \eta_0}A, \qquad B = {\ds 2\eta
\over \ds \eta + \eta_0}A \l{abcsol} \f
The wave impedance of propagating transmitted waves is given by
$\eta = \omega\M_0\M_r/k_{\rm n}$ for TE waves, and $\eta =
k_{\rm n}/(\omega\E_0\E_r)$ for TM waves [$k_{\rm n} = \sqrt{\E_r\M_rk_0^2-k_{\rm t}^2}$ denotes the normal to the interface wave vector component; the formula applies for passive lossy materials with ${\rm Im}\{\E_r, \M_r\} < 0$ (or ${\rm Im}\{\E_r, \M_r\}\rightarrow -0$) if the square root branch is chosen so that ${\rm Im}\{k_{\rm n}\} < 0$ (or ${\rm Im}\{k_{\rm n}\}\rightarrow -0$); the time
dependence is in the form $\exp(+j\o t)$].

In a Veselago
medium both $\E_r$ and $\M_r$ are negative. In the same time,
$k_{\rm n}$ is also negative because the propagating waves are
backward waves. Nothing especially interesting comes out of \r{abcsol} in
this case except that when $\E_r=\M_r=-1$ the interface is
perfectly matched: $\eta = \eta_0$, $C=0$, $B=A$.

But let us suppose that the incident and transmitted waves are
evanescent. Then, $\eta = j\omega\M_0\M_r/\alpha$ for TE waves, and
$\eta = \alpha /(j\omega\E_0\E_r)$ for TM waves. Because the transmitted wave must decay from
the interface, $\alpha$ is positive. We see that for evanescent modes the
ideal case when $\E_r=\M_r=-1$ leads to purely imaginary wave
impedances such that $\eta = -\eta_0$! A resonance occurs:
$C=B\rightarrow \infty$. The reason for such resonant growth of the field
amplitudes when $\E_r, \M_r \rightarrow -1$ is in the excitation
of a surface mode (surface polariton) associated with the
interface. Indeed, if there is no incident field in \r{abc} ($A =
0$) and $\E_r=\M_r=-1$ we can observe that for any $k_{\rm t}
> k_0$ (imaginary wave impedances) there is a solution $C=B \neq
0$ corresponding to a surface wave concentrated near the
interface.

Based on similar considerations several authors\cite{comment,Rao} explained the
evanescent mode amplification in the Pendry lens as the result of
resonant excitation of a pair of coupled surface modes
(polaritons) sitting at the slab interfaces. Under certain
conditions the polariton excited at the second (output) interface
is much stronger than that excited at the first interface.
The exponentially decaying trail of the polariton sitting at the
output interface appears as an exponential growth of the field
inside the slab.

The effects taking place in the material depth (backward waves) and the properties of the
slab interfaces (polariton resonances) both contribute to the Pendry's lens operation. However, it can be shown that in general the presence of a bulk material layer is not crucial. Conceptually, if one can realize a planar sheet such that traveling waves refract negatively when crossing this sheet, a
system of two such sheets placed {\it just in free space} will
focus the propagating modes of a source just like a Veselago slab.
If the sheets also support surface waves for all $k_{\rm t} > k_0$,
then such system will posses surface polariton resonances
reconstructing the evanescent spectrum as well. We found in our
recent paper\cite{ourlens} that a system of two phase conjugating interfaces in
air behaves as a perfect lens. A possible drawback of phase
conjugating design is the necessity to utilize non-linear effects like wave
mixing. In this paper we will discuss alternative
possibilities to evanescent spectrum reconstruction not involving
non-linearity. The design will be based on the principle mentioned
above: We will make use of a couple of polariton-resonant surfaces or grids
placed in free space. No bulk backward-wave materials will be
involved, providing more flexibility and less limitations in
design.

\section{Analysis based on transmission matrices}

In this section and in what follows we restrict our consideration by
the evanescent spectrum only. Our purpose here will be to find such
conditions on resonating sheets that lead to
``amplification'' of the evanescent modes in the proposed
double-grid system. We will call the system simply as {\it
device}. A possible name for such a device can be
{\it near-field lens}, but we would prefer not to use
word {\it lens} in this context to avoid misunderstanding. Let us
emphasize that our aim here is the restoration of the near-field
or evanescent field picture of a source. The systems to be considered in the
following do not focus
propagating modes. This can be done by other well-known optical
means.

We will make use of  a powerful method based on so-called $2\x2$ wave
transmission matrices, well known in the microwave circuit theory.\cite{microwave}
These matrices connect the complex amplitudes of waves traveling
(or decaying) in the opposite directions in a waveguiding system
or a system where one can determine the principal axis of propagation
and measured at two reference planes:
\e \left(\begin{array}{c}
E_2^{-}\\
E_2^{+}\\
\end{array}\right)
=\left(\begin{array}{cc}
t_{11} & t_{12}\\
t_{21} & t_{22}\\
\end{array}\right)
\.
\left(\begin{array}{c}
E_1^{-}\\
E_1^{+}\\
\end{array}\right) \f
Here, $E_{1}^{\pm}$ and $E_{2}^{\pm}$
denote the tangential components of the electric field complex amplitudes
of  waves at
the first (input) and the second (output) interfaces of a device,
respectively (we restrict ourselves by planar layered structures and plane
waves). The signs $^\pm$ correspond to the signs in the propagator
exponents $e^{\pm jk_{\rm n} z}$ of these waves, and $z$ is the
axis orthogonal to the interfaces (the main axis of the system).
It is known that the T-matrix
of a serial connection of several devices described by their
T-matrices is simply a multiplication of the matrices in the order
determined by the connection.

Our purpose is to build a theoretically ideal near-field imaging device. Hence,
the total transmission matrix from the source plane to the plane where
the source field distribution
is reconstructed must be the identity matrix
\e
T_{\rm tot} = T_{\rm sp\ after}\.T_{\rm dev}\.T_{\rm sp\ before} =
\left(\begin{array}{cc}
1 & 0\\
0 & 1\\
\end{array}\right) \l{Ttot} \f
for  every spatial harmonic of the
source field. Here, $T_{\rm sp\ before}$ and $T_{\rm sp\
after}$ represent the air layers occupying the space between the source plane
and the device, and the space between the device and the image plane.
$T_{\rm dev}$ is the transmission matrix of the device.
From this formula it is obvious that
a {\it complete} reconstruction of the field  distribution in the source plane
at a distant image
plane must involve phase compensation for the propagating space
harmonics and ``amplification'' for the evanescent ones. In other
words, one needs to synthesize a device that somehow inverts the action
of a free-space layer. A slab of a material with $\epsilon_r=-1$ and
$\mu_r=-1$ (Veselago medium)\cite{Veselago,Pendry} and a pair of parallel conjugating
surfaces or sheets \cite{ourlens} operate as such device.   In this paper we will find
other {\it linear} solutions working for the evanescent fields of a source.

Let us note here that condition \r{Ttot} is a strict condition requiring not only
the one-way transmission to be such that it reconstructs
the source field picture at the image plane, but also the matching
to be ideal (no reflections from the device) and the device operation to be symmetric (reversible
in the optical sense).
We will consider some less strict conditions later.

Let us suppose that the source and the image planes are distanced by $d/2$ from the input and the output interfaces of the device. A space layer of thickness $d/2$ has the T-matrix
\e T_{\rm sp}(d/2) = \left(\begin{array}{cc}\exp({-jk_{\rm n}d/2}) & 0\\
0 & \exp({+jk_{\rm n}d/2})\\\end{array}\right) \f
To compensate the action of two such layers before and after the device and satisfy the condition \r{Ttot}, the device
T-matrix $T_{\rm dev}$ has to be, obviously, the inverse of the
transmission matrix of these space layers:
\e T_{\rm dev} =
\left(\begin{array}{cc}\exp({+jk_{\rm n}d}) & 0\\ 0 & \exp({-jk_{\rm
n}d})\\\end{array}\right) \l{Tideal} \f

Let us study if a device modeled by this transmission matrix can be
realized as a combination of two
``field transformers" (e.g., thin sheets of certain
electromagnetic properties) separated
by a layer of free space, like it was discussed in the introduction.
This system is modeled by the transmission matrix
$$
T_{\rm dev} = T_{\rm out}\.T_{\rm sp}(d)\.T_{\rm in} =
$$
\e
\left(\begin{array}{cc}
a & b\\
c & d\\
\end{array}\right)\.
\left(\begin{array}{cc}
\exp({-jk_{\rm n}d}) & 0\\
0 & \exp({+jk_{\rm n}d})\\
\end{array}\right)\.
\left(\begin{array}{cc}
e & f\\
g & h\\
\end{array}\right)
\l{great}
\f
Here, $T_{\rm in}$ and $T_{\rm out}$ are matrices with yet unknown components
describing the two sheets or layers forming the device, and $T_{\rm sp}(d)$ is the matrix of the free-space layer between the sheets.
It is easy to show that if
$a=d=0$, $e=h=0$, and $bg=cf=1$, then the total device T-matrix
takes form  \r{Tideal},
i.e., it is the necessary matrix of a perfect lens.
From the mathematical point of view such an amazing result is simply
an effect of permutation of the matrix components under the multiplication \r{great}.
The physical question which we will need to answer later is how to
realize an interface with a T-matrix of the form
\e
T=\left(\begin{array}{cc}
0 & b\\
c & 0\\
\end{array}\right)\l{T0}
\f
We will show that even this ultimate case is realizable (for excitation
by evanescent fields) by passive grids with specific
electric and magnetic susceptibilities.

However, let us consider a couple of simpler systems also,
resulting from the conditions less strict than \r{Ttot}. We keep
the same operation principle defined by \r{great}. If we allow a
mismatch at the device interfaces still maintaining the device
symmetry, the following solution can be found:
\e T_{\rm in} = T_{\rm out} = \left(\begin{array}{cc}
a & b\\
c & 0\\
\end{array}\right)\l{T1} \f
If $b = -c = 1$, we have for the device
\e
T_{\rm dev} = \left(\begin{array}{cc}
a^2\exp({-jk_{\rm n}d}) - \exp({+jk_{\rm n}d}) & a\exp({-jk_{\rm n}d})\\
-a\exp({-jk_{\rm n}d}) & -\exp({-jk_{\rm n}d})\\
\end{array}\right) \l{Tdev1}
\f
which corresponds to the device scattering matrix (well-known S-matrix) of the form
$$
S_{\rm dev} = \left(\begin{array}{cc}
-t_{21}/t_{22} & 1/t_{22}\\
t_{11} - t_{12}t_{21}/t_{22} & t_{12}/t_{22}\\
\end{array}\right) =
$$
\e
-\left(\begin{array}{cc}
a & \exp({+jk_{\rm n}d})\\
\exp({+jk_{\rm n}d}) & a\\
\end{array}\right)
\f
The S-matrix elements are the reflection and transmission
coefficients for two ``ports'' of our device. One can see that the device
``amplifies'' evanescent modes (due to pluses in the $s_{21}$
and $s_{12}$ exponents) and reflects in both ports with the
reflection coefficient equal to $s_{11} = s_{22} = -a$.

If the symmetry is not important but the matching is, the following solution is possible:
\e T_{\rm out} = \left(\begin{array}{cc}
a & b\\
c & 0\\
\end{array}\right), \qquad
T_{\rm in} = \left(\begin{array}{cc}
0 & f\\
g & h\\
\end{array}\right) \l{T2} \f
If $bg=cf=1$, the total T-matrix for the device becomes
\e
T_{\rm dev} = \left(\begin{array}{cc}
\exp({+jk_{\rm n}d}) & af\!\exp({-jk_{\rm n}d})+bh\exp({+jk_{\rm n}d})\\
0 & \exp({-jk_{\rm n}d})\\
\end{array}\right) \l{Tdev2}
\f
which corresponds to the scattering matrix of the form
\e
S_{\rm dev} =
\left(\begin{array}{cc}
0 & \exp({+jk_{\rm n}d})\\
\exp({+jk_{\rm n}d}) &bh\exp({+2jk_{\rm n}d}) + af\\
\end{array}\right)
\f
One can see that the device is matched for the waves coming to the
first interface ($s_{11} = 0$) and also ``amplifies'' the
evanescent modes. In the next section we will describe the ways to
realize matrices \r{T0}, \r{T1}, and \r{T2}.

\section{The use of impedance sheets}

At first we consider a simple system: a lossless isotropic grid, e.g., a
conductive wire mesh (possibly loaded by certain bulk reactances inserted in every cell). If the grid supports only electric currents, and there is no effective magnetic current induced on the grid, then the grid
reflection coefficient $R$ and transmission coefficient $T$  at
the grid plane are related as
\e T = 1 + R \f
provided that they
are defined through the electric field tangential components (for a given polarization). The
corresponding T-matrix of such a grid is
\e T_{\rm g} =
\left(\begin{array}{cc}
\ds {1+2R\over 1+R} & \ds {R\over 1+R}\\[3mm]
\ds -{R\over 1+R} & \ds {1\over 1+R}\\
\end{array}\right)
\f

It is possible to make grids supporting propagation of surface
modes (also known as {\it slow waves} in radio engineering). For wire meshes, for
example, this phenomenon is well investigated. \cite{Kontorovich}
If the tangential component of the wave vector of an incident wave
coincides with the propagation factor of a surface mode, the
surface mode resonance takes place. Obviously, the incident wave
should be evanescent in this case to match with the propagation constant of the
surface mode.
At a surface mode resonance $R\rightarrow\infty$ (for evanescent
modes $R$ is not bounded by $|R| \le 1$). Then, the grid T-matrix
takes the form
\e T_{\rm g} = \left(\begin{array}{cc}
2 & 1\\
-1 & 0\\
\end{array}\right) \l{Tg} \f
which is of the necessary form \r{T1}.

For a better understanding we reformulate the consideration above
in terms of the grid impedance. If the boundary condition on the
grid is given as $E_{\rm t} = Z_{\rm g} J$, where $J$ is the averaged electric surface current density induced on the grid, $E_{\rm t}$ is the averaged tangential electric field in the grid plane,
and $Z_{\rm g}$ is
the grid impedance, the reflection coefficient can be found
as  \cite{modeboo}
\e
R = -\left(1 + {2Z_{\rm g}\over\eta_0}\right)^{-1}
\l{Rzgrid}
\f
and the grid transmission matrix becomes
\e T_{\rm g} =
\left(\begin{array}{cc}
\ds 1-{\eta_0\over 2Z_{\rm g}} & \ds -{\eta_0\over 2Z_{\rm g}}\\[3mm]
\ds {\eta_0\over 2Z_{\rm g}} & \ds 1+{\eta_0\over 2Z_{\rm g}}\\
\end{array}\right)
\l{Tzgrid}
\f

The reflection coefficient \r{Rzgrid} becomes
infinite and the transmission matrix \r{Tzgrid} reduces to \r{Tg}
when $\o$ and $k_{\rm t}$ satisfy equation
\e Z_{\rm g}(\o, k_{\rm t}) + {\eta_0(\o, k_{\rm
t})\over 2} = 0 \l{disp}\f
which is the dispersion equation for
surface modes on the grid surface.
Because
$\eta_0$ is purely imaginary for evanescent modes (it is
inductive for TE waves and capacitive for TM waves) one can see
that in principle there are no restrictions on realizing a
capacitive or inductive grid or array possessing the necessary
resonance for some value(s) of $k_{\rm t}$.

We can see from \r{Tzgrid} that it is enough to change
the sign of the grid impedance to realize the second matrix in
\r{T2}. Such grid is not at resonance with the incident evanescent
field, and it works as an additional matching layer or a
load for the output grid which must experience a strong resonance in
accordance with \r{T2}.

Let us now consider a more complicated grid or array that supports
both electric and magnetic currents. We suppose that the electric
current is excited by electric fields in the array plane and the
magnetic current is due to magnetic fields at the same plane. In
the presence of two currents, the tangential components of  both
electric and magnetic fields are not continuous across the
interface:
\e
E_1-E_2 = J_{\rm m}, \quad H_1 - H_2 = J_{\rm e}
\f
where $J_{\rm e}$ and $J_{\rm m}$ stand for the averaged electric and magnetic surface current densities. The following conditions determine the current amplitudes in terms of two grid impedances $Z_{\rm e}$ and $Z_{\rm m}$:
\e
{E_1 + E_2\over 2} = Z_{\rm e} J_{\rm e}, \quad Z_{\rm m}{H_1 + H_2\over 2} = J_{\rm m}
\f
It is possible to show that an interface defined by the above conditions has the following T-matrix:
$$
T_{\rm g} = \Bigg[{4Z_{\rm e}\over Z_{\rm m}} - 1\Bigg]^{-1}\x
$$
\e
\left(\begin{array}{cc}
\ds \Bigg[1-{2Z_{\rm e}\over \eta_0}\Bigg]\Bigg[1-{2\eta_0\over Z_{\rm m}}\Bigg] &
\ds 2\Bigg[{Z_{\rm e}\over \eta_0} - {\eta_0\over Z_{\rm m}}\Bigg]\\[3mm]
\ds -2\Bigg[{Z_{\rm e}\over \eta_0} - {\eta_0\over Z_{\rm m}}\Bigg]&
\ds \Bigg[1+{2Z_{\rm e}\over \eta_0}\Bigg]\Bigg[1+{2\eta_0\over Z_{\rm m}}\Bigg]\\
\end{array}\right)
\l{Tsupergrid}
\f
Eq. \r{Tzgrid} is a particular case of \r{Tsupergrid} when $Z_{\rm
m}$ tends to zero and the magnetic current vanishes. The matrix
\r{Tsupergrid} reduces to form \r{T0} in two cases.  First, this
happens when the electric subsystem is at resonance: $Z_{\rm e} =
-\eta_0/2$ and the magnetic subsystem works as a loading: $Z_{\rm
m} = 2\eta_0$. In this case
\e
T_{\rm g} = \left(\begin{array}{cc}
0 & 1\\
-1 & 0\\
\end{array}\right)
\f
Or, second, when the roles are interchanged:
$Z_{\rm e} = \eta_0/2$, $Z_{\rm m} = -2\eta_0$. For this case
\e
T_{\rm g} = \left(\begin{array}{cc}
0 & -1\\
1 & 0\\
\end{array}\right)
\f
Let us note again that nothing forbids
realization of the necessary impedances for a given
$k_{\rm t}$,  because for  the evanescent modes they are purely imaginary.

\section{Arrays of weakly interacting resonant inclusions}

In the previous sections we have shown that the surface mode
resonance plays the key role in the mechanism of the evanescent
field amplification. We also have proven that such resonance is
possible to realize in a passive grid or array. The resonance
occurs when the~$(\o, k_{\rm t})$ pair belongs to the
polariton spectrum of the grid.

Practically speaking, this means that for a given frequency $\omega$ one may
realize one or at most several resonant values of
the transverse propagation constant $k_{\rm t}$.
It can be enough for some purposes as, for example, for
resonant extraction and ``amplification'' of a certain spatial harmonic of the incident
field. However, for a device operating as a near
field lens one should provide as wide range of operable $k_{\rm
t}$ as possible.

Mathematically (and ideally), the last means that the dispersion
equation \r{disp} should be somehow turned at a given frequency
into an identity for any $k_{\rm t} > k_0$. Although that is
impossible in practice, there is a good approximation for this. Let
us consider a dense regular two-dimensional array of small resonant
dipole inclusions. Instead of writing the boundary
condition in terms of the {\it total} averaged field in the array
plane as before, we may solve the excitation problem directly in terms of the
{\it external} field and the induced dipole moments. In a given external
field $E_{\rm ext}$, the dipole moment of each particle in the array is\cite{modeboo}
\e
p = \chi(\o) \Big[E_{\rm ext} + \beta(\o, k_{\rm t})p\Big]
\f
Here $\chi$ is the particle polarizability,
and $\beta(\o, k_{\rm t})$ is so-called interaction
factor, which is a function of $k_{\rm t}$. Obviously, the solution
for the induced dipole moment is
\e
p = {\ds E_{\rm ext}\over \ds {1\over \chi(\o)} - \beta(\o, k_{\rm t})}
\l{moment}
\f
From here it is already seen what condition is needed for a
resonance, but let us proceed a bit further. In terms of the
average surface current density $J = {j\o p/S_0}$ ($S_0$ is the unit cell area) we write
\e
Z_{\rm c}J = E_{\rm ext},
\f
where
\e
Z_{\rm c} = {S_0\over j\o}\left[{1\over \chi(\o)} - \beta(\o, k_{\rm t})\right]
\l{cellimp}
\f
Here we have introduced a quantity which we call {\em cell impedance}
$Z_{\rm c}$. This impedance is related with the grid impedance.
Indeed, because the total tangential electric field at the array
plane is $E = E_{\rm ext} - \eta_0 J/2$, we have
\e
Z_{\rm g} = Z_{\rm c} - {\eta_0\over 2}
\f
Comparing the last relation with the resonance condition~\r{disp}
we see that the polariton resonance takes place when $Z_{\rm c} =
0$. Of course, the same conclusion follows directly from looking
at the denominator of \r{moment}.

The imaginary parts of $1/\chi$ and $\beta$ cancel out in a lossless
non-radiating array\cite{modeboo} (we work with evanescent fields).
In order to resonate at all harmonics of the evanescent field spectrum,
the resonant condition $Z_{\rm c}=0$ must be satisfied for all
$k_{\rm t}$. Because the inclusions are assumed to be small dipole
particles, their polarizability $\chi$ depends on the
frequency, but does not depend on $k_{\rm t}$. On the other hand,
the interaction constant $\beta$ depends both on the frequency and on
the transverse wave number. Thus, the only possibility to
realize such grids using small inclusions is to use resonant
particles ($\Re\{1/\chi\}\rightarrow 0$) and minimize the
field interactions between
the particles in the array (provide
$\Re\beta(\o, k_{\rm t})  \rightarrow 0$).
If these conditions are satisfied, each particle in the grid
is excited {\em locally} by the incident field at its position
(if the field interaction in the array is negligible, the local
field equals the incident field). Since at the operational frequency
the particles are at resonance, {\em arbitrary} spatial distributions of
the incident field will excite a resonance of the whole array.
In other words, at this frequency the grid indeed supports polaritons with
arbitrary $k_{\rm t}>k_0$, as needed for evanescent field amplification and
imaging.

In a real system particles always
interact, and the last condition cannot be exactly
satisfied. However, it is possible to reduce or compensate interactions at a
certain frequency.
In the next
section we will describe an experiment based on
implementing an array of highly resonant but weakly interacting
particles. The experiment will demonstrate
evanescent field enhancement in a passive linear system.
Another possibility is to use inclusions of larger sizes, and
try to compensate the spatial dispersion of the particle interaction
with the spatial dispersion of the response of one inclusion.
We will not explore this last possibility in this paper.

\section{Experiment}

\begin{figure}[ht]
\centering \epsfig{file=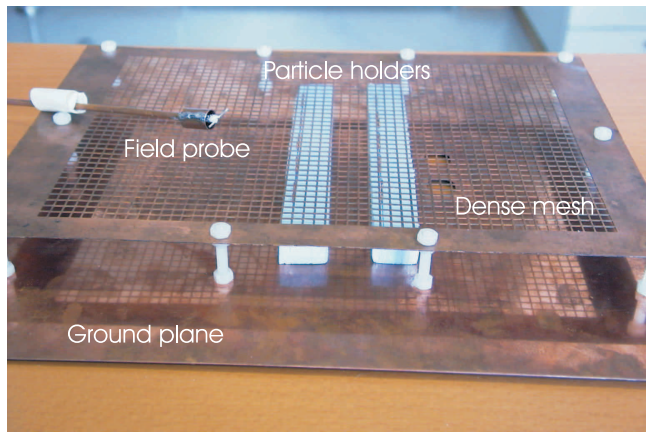, width=8.5cm}
\caption{Photo of the experimental set-up.
Resonant particles were positioned on two parallel foam holders
located between two highly conducting planes. The
probe used to scan the field distribution is seen on the top.
A similiar antenna was used as the source (Not shown on the
photo. It was
positioned between the conducting planes.)} \label{photo}
\end{figure}

In the microwave experiment (the operating frequency was close to
5 GHz),
evanescent fields were generated in the space between a
metal plate and a dense mesh of conducting strips forming a two-plate
waveguide, see
Figure~\ref{photo}. Microwave absorbers were used around the system
to minimize reflections from the open ends of the waveguide. The upper
screen was made weakly penetrable to the fields in order to give us a possibility
to measure the field distribution by a probe positioned on top of the
mesh. The transmission coefficient of this mesh (for normal plane-wave
incidence) at the operation frequency was about $-20$ dB.
As a source, we used a wire dipole antenna whose length was close to
$\lambda/4$. The dipole was parallel to the conducting plates, so that
only evanescent modes were excited in the space between the plates
(the distance between the plates was $2.5$ cm, so that all the waveguide modes of this
polarization were evanescent).

\begin{figure}[ht]
\centering \epsfig{file=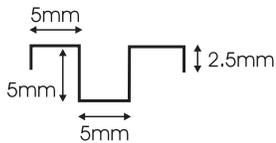, width=0.2\textwidth}
\caption{Small resonant particle.} \label{one}
\end{figure}

As was established above, to realize a device that would ``amplify'' evanescent fields
we need to design an array of small resonant particles that weakly interact.
To validate this concept, one can minimize interactions between
particles simply increasing the distance between the particles in the
array. In our first experiment, we measured fields in a system of only {\em two}
resonant particles, which corresponds to the case of two parallel
arrays with infinitely large periods.
The particles were made of a copper wire of $0.8$ mm diameter, and their shape and
dimensions were as shown in Figure~\ref{one}.
The wire was meandered in order to make the overall dimensions small
as compared with the wavelength. The stretched wire length was close to
$\lambda/2$, so the particles showed resonant response.

\begin{figure}[ht]
\centering \epsfig{file=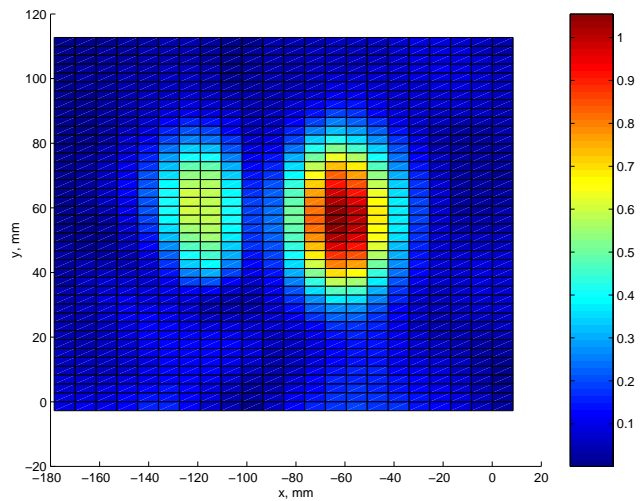, width=8.5cm}
\caption{The distribution of evanescent
field created by a small dipole antenna in the presence of two small
resonant particles. The source dipole is placed at $x=-125$ mm, $y=55$ mm and directed along the $y$-axis. Two metal particles (Figure~\ref{one}) are placed at $x=-105$ mm, $y=55$ mm and $x=-65$ mm, $y=55$ mm. The particles are oriented along the $y$-axis. The frequency is 5 GHz. The probe is $0.5$ cm away from the top mesh of the setup. The field amplitude scale is linear.} \label{one_scan}
\end{figure}
\begin{figure}[ht]
\centering \epsfig{file=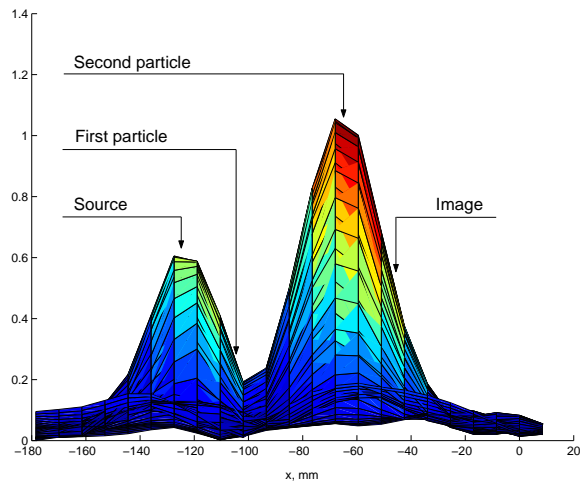, width=8.5cm}
\caption{Dependence of the field amplitude along the device axis for
the same arrangement as in Figure~\ref{one_scan}. This is a side view of the 3D-plot of the field distribution shown in Figure~\ref{one_scan}. The key positions on the plot are indicated by arrows.} \label{one_profile}
\end{figure}

The experimental results are shown in Figures~\ref{one_scan} and
\ref{one_profile}. It can be seen that, as expected from the
theory, the first particle is very weakly excited, and a high-amplitude
plasmon polariton is sitting at the second particle. The field amplitude in the
``image plane'' is close to that at the source position.

Although this experiment demonstrates the validity of the main principle
of near-field enhancement and a possibility to restore the
evanescent field components, grids with reasonably
small periods are necessary to realize an imaging device.

\begin{figure}[ht]
\centering \epsfig{file=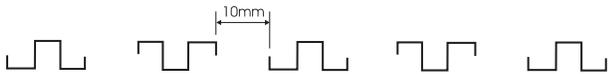, width=8cm}
\caption{A periodic array of small resonant particles.} \label{array_geom}
\end{figure}

\begin{figure}[ht]
\centering \epsfig{file=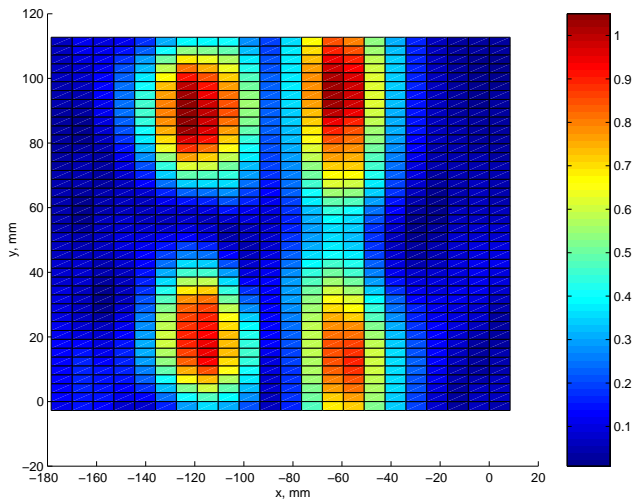, width=8.5cm}
\caption{The distribution of evanescent
field created by two small dipole antennas in the presence of two grids of small
resonant particles. The field complex values are measured at $5.15$ GHz (1st
polariton resonance) and $5.26$ GHz (2nd polariton resonance) and summed up.
Two source dipoles are placed at $x=-125$ mm, $y=20$ mm and at $x=-125$ mm, $y=90$ mm. The dipoles are oriented along the $y$-axis.
Two grids (5 particles in each) are placed at $x=-105$ mm, $y=55$ mm and $x=-65$ mm,
$y=55$ mm along the $y$-axis. Probe is $0.5$ cm away from the top mesh of the setup. The field scale is linear.} \label{two}
\end{figure}

To study phenomena in such grids,
we have made measurements in a system of two regular arrays of
similar particles. The array geometry is shown in Figure~\ref{array_geom}.
In this system, the field interaction of particles exists,
meaning that maintaining polariton resonance for all
transverse wave numbers is not possible.
In the measurements, we first experimentally determined the eigenfrequencies of the
grids. Each eigenfrequency corresponds to a certain transverse
wavenumber $k_{\rm t}$. Next, we exited the grids at each of these
frequencies (2 frequencies were practically used) and superimposed  the
measured spatial profiles. This corresponds to reconstructing the
source spatial spectrum using only a few spatial harmonics.
The result for the case of excitation by two small dipole
antennas is shown in Figure~\ref{two}.
Strong excitation of the second grid is clearly visible, as well as
an image of the source field behind the grids. This last experiment
should be considered as a first step only, because
no effective reduction of the field interactions between the grid particles
was realized, and  the set-up had many non-idealities. However,
we can conclude that the experiments successfully
validate the principle of near-field enhancement in simple
passive and linear resonant systems.

\section{Conclusions}

In this paper we have considered a wide class of passive linear structures able to
enhance evanescent fields and reconstruct the near-field image
of a source. All these structures result from the idea of
using a system of two parallel polariton-resonant grids or arrays
separated by a certain distance and placed in free space.

The physics behind this idea is based on the known behavior of coupled resonant systems. If in a system of two resonators the first resonator is pumped by an external force and the second resonator is coupled to the first one, then under certain conditions it is possible for the amplitude of oscillations in the second resonator to be much higher than the amplitude of the external field and the amplitude of the first resonator oscillations. A similar interpretation of the phenomena taking place in a coupled-polariton-resonant system (the Veselago slab) can be found in a recent work by Rao and Ong.\cite{Rao}

The present paper continues of the research we started in order to eliminate the need in backward-wave or other exotic {\it bulk material layers} in the design of near-filed imaging devices. In our recent work\cite{ourlens} we showed that a system of two phase-conjugating planes or sheets placed in free space 
behaves as a perfect lens proposed by Pendry.\cite{Pendry} The obvious drawback of the phase-conjugating design is the necessity to involve nonlinear materials or devices in the structures realizing the conjugating sheets. Here we have shown that if the focusing of the propagating spectrum is not required there exist several {\it linear} solutions. 

We have developed a general synthesis approach based on the wave transmission matrices\cite{microwave} to find the conditions under which a system of two coupled polariton-resonant grids or arrays enhances incident evanescent field. Next, the inner design of the grids has been revealed, and it has been shown that there are many possibilities arising from the use of impedance sheets. The necessary structures can be realized as arrays of weakly interacting resonant particles of different nature.

We have experimentally confirmed the possibility to use passive linear polariton-resonant systems for evanescent field enhancement at microwaves. The experiment supports the main concepts of our theoretical findings. The resonant growth of the evanescent fields coming through the system has been observed.

\section*{Acknowledgment}
This work has been partially supported by the Academy of Finland and TEKES through the Center-of-Excellence program.

\end{document}